\documentclass[prl,twocolumn,aps,superscriptaddress,preprintnumbers]{revtex4}

\usepackage{amsmath}
\usepackage{amsfonts}
\usepackage{hyperref}

\newcommand{\be}{\begin{equation}}
\newcommand{\ee}{\end{equation}}
\newcommand{\bea}{\begin{eqnarray}}
\newcommand{\eea}{\end{eqnarray}}
\newcommand{\beq}{\begin{equation}}
\newcommand{\eeq}{\end{equation}}
\newcommand{\beqn}{\begin{eqnarray}}

\newcommand{\eeqn}{\end{eqnarray}}
\newcommand{\nn}{\nonumber}

\newcommand{\fft}[2]{\frac{#1}{#2}}
\newcommand{\ft}[2]{{\textstyle\frac{#1}{#2}}}

\def \bege {\begin{equation}}
\def \ende {\end{equation}}
\def \beges {\begin{eqnarray}}
\def \endes {\end{eqnarray}}

\begin{document}
\preprint{LCTP-17-02}

\title{A One-loop Test of Quantum Black Holes in Anti de Sitter Space}

\author{James T. Liu }\email{jimliu@umich.edu}
\affiliation{Leinweber Center for Theoretical Physics, University of Michigan, Ann Arbor, MI 48109, USA}

\author{Leopoldo A. \surname{Pando Zayas}}\email{lpandoz@umich.edu}
\affiliation{Leinweber Center for Theoretical Physics, University of Michigan, Ann Arbor, MI 48109, USA}
\affiliation{The Abdus Salam International Centre for Theoretical Physics, Strada Costiera 11, 34014 Trieste, Italy}

\author{Vimal Rathee }\email{vimalr@umich.edu}
\affiliation{Leinweber Center for Theoretical Physics, University of Michigan, Ann Arbor, MI 48109, USA}

\author{Wenli Zhao}\email{wz10@princeton.edu}
\affiliation{Department of Physics, Princeton University, Princeton, NJ 08544, USA}

\begin{abstract}
Within eleven dimensional supergravity we compute the logarithmic correction to the entropy of magnetically charged asymptotically AdS${}_4$ black holes with arbitrary horizon topology. We find perfect agreement with the expected microscopic result arising from the dual field theory computation of the topologically twisted index. Our result relies crucially on a particular limit to the extremal black hole case and clarifies some aspects of quantum corrections in asymptotically AdS spacetimes.
\end{abstract}

\maketitle


{\bf Introduction:} The Bekenstein-Hawking entropy of a black hole is proportional to the area of its event horizon: $S=k_B c^3 A/(4G_N \hbar) $. Given the fundamental constants involved, its complete understanding necessarily involves  thermodynamical, relativistic, gravitational, and quantum aspects.  Studying corrections to the Bekenstein-Hawking entropy is crucial for a  full understanding of the microscopic degrees of freedom responsible for the macroscopic entropy. In this letter we report on a computation of the one-loop effective action for a class of asymptotically AdS${}_4$ black holes that matches precisely the coefficient of the logarithmic correction arising from a microscopic description. 

The framework for our computation is the Anti de Sitter/Conformal Field Theory (AdS/CFT) correspondence which conjectures the mathematical equivalence of string theory (containing gravity) in asymptotically AdS spacetimes and certain conformal field theories.  It provides, by construction, a non-perturbative definition of quantum gravity in asymptotically AdS spacetimes which is capable, in principle, of addressing puzzling questions of black holes using field theory techniques. Only recently, however, has an explicit example in AdS$_4$/CFT$_3$ emerged. It has been shown that  in the
large-$N$ limit the topologically twisted index of a certain Chern-Simons theory coupled to matter, known as the  ABJM theory, correctly reproduces the leading term in the entropy of magnetically charged black holes in asymptotically AdS$_4$ spacetimes  \cite{Benini:2015eyy}.   Similar matches have now been established in various other situations including: dyonic black holes \cite{Benini:2016rke}, black holes with hyperbolic horizons   \cite{Cabo-Bizet:2017jsl}, and black holes in massive IIA theory  \cite{Azzurli:2017kxo,Benini:2017oxt, Hosseini:2017fjo}. 

Having established the microscopic counting, it is natural to embark on an exploration of the  sub-leading in $N$ structure. In our previous work we studied corrections to the topologically twisted index using a combination of numerical and analytical techniques and identified a logarithmic correction of the form $-\frac{1}{2}\log N$ \cite{Liu:2017vll}.  A corresponding computation on the gravity side, focusing on the near horizon contribution to the one-loop effective action and following the quantum entropy formalism developed by Sen \cite{Sen:2008yk,Sen:2008vm}, however, failed to match this microscopic result \cite{Liu:2017vll,Jeon:2017aif}.  However, here we report that perfect agreement is achieved when the one-loop supergravity computation is performed in the full AdS$_4$ black hole background, and not just in the near horizon geometry.  This suggests that, in contrast with asymptotically flat black holes, the microscopic degrees of freedom of AdS black holes are sensitive to the background in which they are embedded.


{\bf The topologically twisted index in  ABJM theory:}
On the microscopic side, the CFT dual to magnetically charged AdS$_4$ black holes is given by ABJM theory with background flavor fluxes turned on.
ABJM theory is a three-dimensional Chern-Simons-matter theory with $U(N)_k\times U(N)_{-k}$ gauge group
and opposite integer levels $k$ and $-k$ \cite{Aharony:2008ug}. The matter sector contains four complex scalar fields $C_I, (I=1,2,3,4)$ in the bifundamental representation $({\bf N}, \bar{\bf N})$, together with their fermionic partners. The theory  is superconformal and has $\mathcal{N}=6$ supersymmetry generically, but for $k=1,2$, the symmetry is enhanced to $\mathcal{N}=8$. Holographically, ABJM describes a stack of $N$ M2-branes probing a $\mathbb{C}^4/\mathbb{Z}_k$ singularity, whose low energy dynamics are effectively described by 11 dimensional supergravity. 

The presence of background fluxes implements a partial topological twist, and is crucial for preserving supersymmetry when the theory is defined on $\Sigma_g\times S^1$, where $\Sigma_g$ is a genus-$g$ Riemann surface corresponding to the horizon topology of the black hole.
The topologically twisted index is then defined as the supersymmetric partition function of the twisted theory, $Z(n_a, \Delta_a)={\rm Tr}\, (-1)^F e^{-\beta H} e^{iJ_a \Delta_a}$.  It depends on the fluxes, $n_a$, through $H$ and on the chemical potentials $\Delta_a$. This index was constructed in \cite{Benini:2015noa} for $\mathcal{N}\geq 2$ supersymmetric theories on $S^2\times S^1$ and computed via supersymmetric localization.  It was then applied to ABJM theory in \cite{Benini:2015eyy}, and evaluated in the large-$N$ limit.

In the large-$N$ limit, and at genus zero, the $k=1$ index takes the form
\begin{align}
F&=-\frac{N^{3/2}}{3}\sqrt{2\Delta_1\Delta_2\Delta_3\Delta_4}\sum\limits_a\frac{n_a}{\Delta_a}
+ N^{1/2}f_1(\Delta_a,n_a) \nonumber \\
&\quad-\fft12\log N+f_3(\Delta_a,n_a)+\mathcal O(N^{-1/2}),
\label{eq:tti}
\end{align}
where $F=\mbox{Re}\log Z$.  The leading $\mathcal O(N^{3/2})$ term was obtained in \cite{Benini:2015eyy}, and exactly reproduces the Bekenstein-Hawking entropy of a family of extremal  AdS$_4$ magnetic black holes admitting an explicit embedding into 11d supergravity \cite{Cvetic:1999xp}, once extremized with respect to the flavor and $R$-symmetries.  The $\mathcal O(N^{1/2})$ term can be identified with $\mathcal O(\alpha'^3R^4)$ corrections in the supergravity, and does not appear to have a simple form.  On the other hand, the $-\fft12\log N$ term, obtained numerically in \cite{Liu:2017vll}, appears to be universal, and is what we wish to reproduce from the gravity side.

In fact, the topologically twisted index can be defined on Riemann surfaces with arbitrary genus \cite{Benini:2016hjo,Closset:2016arn}, and there is a simple relation between the index on $\Sigma_g\times S^1$ and that on $S^2\times S^1$:
$F_{\Sigma_g\times S^1}(n_a, \Delta_a)=(1-g)F_{S^2\times S^1}(\frac{n_a}{1-g},\Delta_a)$. Since  the coefficient of the logarithmic term in  $F_{S^2\times S^1}$ does not depend on $n_a$ we simply have 
\be
F_{\Sigma_g\times S^1}(n_a,\Delta_a)=\dots -\frac{1-g}{2}\log N+\cdots .
\label{Eq:Index_g}
\ee
We now demonstrate that this logarithmic correction naturally appears in the quantum correction to the extremal magnetically charged  AdS$_4$ black hole.

{\bf One-loop quantum supergravity:}
Since the AdS$_4$ black holes may be embedded in 11d supergravity \cite{Cvetic:1999xp}, we will take a 11d approach to the gravity calculation.
Dimensional analysis shows that logarithmic corrections come from one-loop determinants. The standard computation of such terms for black holes in asymptotically flat spacetime reduces to the  near horizon geometry \cite{Sen:2008vm}. However, in \cite{Liu:2017vll,Jeon:2017aif}, the near horizon contribution  was shown to be $-2\log N$, resulting in a mismatch with the field theory answer. Such a mismatch indicates that either somehow the near horizon geometry is not enough to compute the quantum entropy, or the index does not correctly count microstates in the sub-leading order.

In this letter, we provide evidence for the first possibility by directly computing the logarithmic correction to the entropy from its thermodynamical definition, 
\be
S=\lim_{\beta\to \infty}(1-\beta\partial_{\beta})\log Z[\beta, \dots], 
\label{eq:entropy}
\ee
where $\beta$ is the inverse temperature.  We work in the large AdS radius limit, $L\gg1$, where $L \sim  N^{\frac{1}{6}}$ by the AdS/CFT dictionary.  Our focus is on the one-loop partition function, which can be written schematically as
\be
Z_{\text{1-loop}}[\beta,\ldots]=\sum_{D}(-1)^D(\ft{1}{2}\log \text{det}'D)+\Delta F_0,
\label{entropy1lp}
\ee
where $D$ stands for kinetic operators corresponding to various fluctuating fields and $(-1)^D=-1$ for bosons and $1$ for fermions.  The prime indicates removal of the zero modes, which are accounted for separately by
\be
\Delta F_0=\log \int [d\phi]|_{D\phi=0},
\label{zm}
\ee
where $\exp(-\int d^dx \sqrt{g}\phi D\phi)=1$.

For a stationary background, the logarithmic part of the one-loop determinant comes from 
\begin{equation}
\label{oneloop}
-\ft{1}{2}\log \text{det}'D=\left(\frac{1}{(4\pi)^{\frac{d}{2}}} \int_0^{\beta}dt A_{d/2}(\beta,\dots)-n_0\right)\log L+\cdots,
\end{equation}
where $A_{d/2}(\beta,\dots)=\int d^{d-1}x\sqrt{g}\,a_{d/2}(x,x)$. For odd dimensional spacetimes, the Seeley-De Witt coefficient $a_{\frac{d}{2}}(x,x)$ vanishes due to the lack of a diffeomorphism invariant scalar function of the metric with scaling dimension $d$ \cite{Vassilevich:2003xt}. The advantage of working in 11d is then clear, as only the zero mode contributions remain. The structure of the logarithmic term is then given by
\begin{equation}\label{logterm}
\log Z[\beta, \dots]=\sum_{\{D\}}(-1)^D(\beta_D-1)n_D^0\log L+ \Delta F_{\mathrm{Ghost}}+\cdots,
\end{equation}
where the ghost contributions are treated separately, as in \cite{Bhattacharyya:2012ye}, and $\beta_D$ is due to the integration over zero modes, Eq.~(\ref{zm}), in the path integral, as studied in various cases of logarithmic contributions to the black hole entropy and the one-loop partition function \cite{Sen:2011ba, Banerjee:2010qc, Banerjee:2011jp, Bhattacharyya:2012ye}. 

{\it Magnetically charged AdS$_4$ black holes:}
Our task at hand is thus to enumerate the zero modes of the fluctuations in the AdS$_4$ magnetic black hole background.  These black holes were originally obtained in \cite{Cacciatori:2009iz}, more recently discussed in \cite{Hristov:2010ri} and reviewed in \cite{Benini:2015eyy}. They are solutions of $\mathcal{N}=2$ gauged supergravity with 3 vector multiplets, and with prepotential and FI gauging parameters 
\be
F=-2i\sqrt{X^0X^1X^2 X^3},\qquad \xi_{\Lambda}=\frac{1}{2},\qquad \Lambda=1,\dots,4.
\ee
The family of black holes admits background fluxes $F^a$, $a=1, \dots, 4$ over a Riemann surface horizon $\Sigma_g$. The BPS condition requires 
\be \label{BPSeq}
\frac{1}{2\pi}\sum_a\int_{\Sigma_g}F^a=\chi(\Sigma_g).
\ee
The solutions are parametrized by four fluxes $n^a$ and the genus of the horizon, $g$, subject to the above BPS constraint. The metric of the solution can be put in the form
\be
ds^2 = U^2 (r) \: d \tau^2 + U^{-2} (r) \:dr^2 + h^2 (r) ds^2_{\Sigma_g},
\label{eq:AdSBHg}
\ee
where $U(r)=e^{K(r)}r^2(1-\frac{a}{2gr^2})^2$ and $h(r)= 2e^{K(r)} r^2$ in the extremal case. 
A  more comprehensive review, including non-extremal solutions, is found in  \cite{Chow:2013gba}.

These black holes may be uplifted as solutions to 11d supergravity, with fields consisting of a metric $g_{\mu\nu}$, a three-form field $C_{\mu\nu\rho}$ and a gravitino $\Psi_\mu$.  From an 11d perspective, we are interested in their zero mode fluctuations on a background which is locally of the form $M_4\times S^7$, where $M_4$ has metric given by Eq.~(\ref{eq:AdSBHg}), and the 7-sphere is squashed in the process of turning on magnetic flux. Given an 11d  kinetic operator, one can decompose it to a $M_4$ part and a $S^7$ part. Since compactness of $S^7$ leads to non-negative eigenvalues, zero modes of the 11d supergravity fields are thus simultaneously zero modes in $M_4$ and $S^7$.  As a result, we only need to consider the massless Kaluza-Klein sector, corresponding to the fields of 4d $\mathcal N=8$ gauged supergravity, and to seek out their zero modes in the AdS$_4$ black hole background.

{\it Metric and fermion zero modes:}
From a four-dimensional perspective, the fluctuating fields we must consider include the metric, $p$-forms, and fermions.  We first demonstrate that the metric and fermions do not have any zero modes in the black hole background.  This leaves the $p$-forms, we we turn to below.  For the metric, a zero mode requires a pure gauge mode with a non-normalizable gauge parameter. To show it cannot exist, it is enough to focus on the asymptotic metric, 
\be
ds^2=\frac{dr^2}{r^2}+r^2(dt^2+ds^2_{\Sigma_g}).
\label{eq:metric}
\ee
For a pure gauge deformation, $h_{\mu\nu}=\nabla_{\mu}\eta_{\nu}+\nabla_{\nu}\eta_{\mu}$, normalizability demands
\begin{align}
    h_{rr}&=2\nabla_r\eta_r\kern2.7em\sim {1}/{r^4},\nn\\
    h_{ri}&=\nabla_r\eta_i+\nabla_i\eta_r\sim {1}/{r^2},\nn\\
    h_{ij}&=\nabla_i\eta_j+\nabla_j\eta_i\sim \mathcal{O}(1).
\end{align}
Thus asymptotically $\eta_i\sim1/r$ and $\eta_r\sim1/r^3$.  As a result
\begin{equation}
\|\eta\|^2=\int \sqrt{g}g^{\mu\nu}\eta_{\mu}\eta_{\nu}d^4x\sim \int^\infty (r^4\eta_r^2+\eta_i^2)dr<\infty,
\end{equation}
and the gauge parameter is thus normalizable.

A similar argument can be made for the gravitino to show the absence of zero modes.  In particular, potential gravitino zero modes correspond to would be pure gauge modes $\psi_\mu=\mathcal D_\mu\epsilon$ (where $\mathcal D_\mu$ is the supercovariant derivative), however with non-normalizable spinor $\epsilon$.  Working with the metric (\ref{eq:metric}), we can see that $\epsilon\sim1/r^2$ is required for $\psi_\mu$ to be normalizable.  Since this makes $\epsilon$ normalizable as well, we conclude that there are no gravitino zero modes in this background.


{\it $p$-form zero modes:}
We now turn to an examination of $p$-form zero modes.  Recall that, for zero modes of $A_p$ in a compact space, one requires $\langle dA_p, dA_p\rangle=0$ with respect to the standard inner product on $p$-forms. This amounts to requiring $A_p$ to be closed. But $A_p$ and $A_p+d\alpha_{p-1}$ are gauge equivalent, and the redundant contributions in the path integral are canceled by the Faddeev-Popov procedure. Therefore the number of the zero modes is the dimension of the $p$-th de-Rham cohomology.

We are of course interested in a non-compact space, in which case there are several complications, especially with infinite volume. First, the physical spectrum only includes forms with finite action, as the weight in the Euclidean path integral is $e^{-S}$. Second, for a non-normalizable $p-1$ form, the gauge transformation $d\alpha_{p-1}$ can be normalizable and included in the physical spectrum, yet the Faddeev-Popov procedure can only cancel gauge transformations with normalizable $\alpha_{p-1}$. The result is a physical spectrum with some pure gauge modes with non-normalizable gauge parameters, a situation which is ubiquitous  in one-loop gravity computations in AdS  \cite{Sen:2011ba,Banerjee:2011jp}. Third, there are usually infinitely many such modes, making the number of zero modes infinite. Mathematically, the first two complications lead one to consider $L^2$ cohomology, $H_{L^2}^p(M,\mathbb{R})$ by replacing the de-Rham chain complex by one consisting of $L^2$ $p$-forms whose exterior derivative is also $L^2$ \cite{10.2307/2042193}. The third complication simply states that $\text{dim}\,H_{L^2}^p(M,\mathbb{R})$ can be unbounded. 

A further subtlety in the non-compact case is the difference between $H_{L^2}^p(M,\mathbb{R})$ and, $\mathcal{H}^p_{L^2}(M,\mathbb{R})$, the space of $L^2$ harmonic $p$-forms. As in \cite{CAMPORESI199457}, a transverse condition on the gauge field is imposed when heat kernel method is applied. It is, therefore, more precise to identify the space of concern to be $\mathcal{H}^p_{L^2}(M,\mathbb{R})$. The number $n_p^0$ of $p$-form zero modes is then given by the regularized dimension 
\be
n_p^0=\text{dim}^R\mathcal{H}^p_{L^2}(M,\mathbb{R})=\int_R \sum_n A_p^n \wedge \star A_p^n,
\ee
where $\{A^n_p\}$ is a set of orthonormal basis functions, and the integral is defined as the finite piece after regularization.

Before turning to a full accounting of zero modes, we make an observation that will prove useful below. When the manifold is compact the Euler characteristic is given by
$
\chi(M)=\sum_p(-1)^p\text{dim}\mathcal{H}^p(M,\mathbb{R}),
$
and a similar relation still holds for non-compact manifolds in the class known as conformally compact manifolds (see Corollary 8.1 in \cite{ALBIN20071}). A conformally compact manifold is a manifold with boundary whose metric admits expansions near the boundary 
\be
ds^2=\frac{du^2}{\alpha(u)^2 u^2}+\frac{h_{ij}dx^idx^j}{u^2},
\ee
where the boundary is at $u=0$, with $\alpha(0)\neq 0$ and $h_{ij}(0)$ well defined. For such a manifold of even dimension it was proved in  \cite{ALBIN20071}  that $\mathcal{H}_{L^2}^i=H^k_{DR}(M,\partial M)$ for $i< \frac{n}{2}$ and  $\mathcal{H}_{L^2}^i=H^k_{DR}(M)$ for $i>\frac{n}{2}$. The appropriate modification of the Gauss-Bonnet theorem states 
\begin{align}
\int^{\mathrm{Reg}} \text{Pf}(R)&=2\sum_{i<\frac{n}{2}}(-1)^i\text{dim}H^i_{DR}(M,\partial M)\nonumber\\
&\quad+(-1)^{\frac{n}{2}}\text{dim}^R\mathcal{H}_{L^2}^{\frac{n}{2}}(M,\mathbb{R}),
\label{keyeq}
\end{align}
where $H^i_{DR}(M,\partial M)$ stands for the relative de-Rham cohomology, and the Gauss-Bonnet integral is regularized. It follows from the definition that an asymptotic AdS manifold is a conformally compact manifold and  Eq.~(\ref{keyeq}) applies to determine  $\text{dim}^R\mathcal{H}_{L^2}^{\frac{n}{2}}(M,\mathbb{R})$ for the AdS$_4$ black hole. Indeed, an explicit  version of the above formula was applied in \cite{Larsen:2015aia} to elucidate aspects of quantum inequivalence in AdS$_4$.

In applying the thermodynamic entropy (\ref{eq:entropy}), we take the extremal limit of the non-extremal AdS$_4$ black hole.  In this case, the topology of the non-extremal black hole is homotopic to its horizon $\Sigma_g$ due to the contractible $(t,r)$ directions. Thus the Euler characteristic of the non-extremal black hole is simply $\chi_{\text{BH}}=2(1-g)$. It also indicates that all but the second relative de-Rham cohomology vanish. Therefore, using Eq.~(\ref{keyeq}), one obtains 
\be
n_2^0=\text{dim}^R\mathcal{H}_{L^2}^{2}(M,\mathbb{R})=\int^{\text{Reg}} \text{Pf}(R)=\chi_{\text{BH}}=2(1-g),
\label{2form}
\ee
and moreover these are the only possible zero modes in the black hole background.

The regularized dimension, $n_2^0$, can be negative for higher genus. In fact, this is a general feature of regularized dimensions defined as above. For example, in the case of AdS$_2$, $\text{dim}^R\mathcal{H}_{L^2}^{1}(\text{AdS}_2,\mathbb{R})=-1$ and such negative dimensions occurs in various computations of the macroscopic logarithmic contributions to BPS black holes in asymptotically flat spacetime \cite{Banerjee:2010qc, Banerjee:2011jp}. 

{\it Two-form zero modes from 11d supergravity:} What we have seen above is that the logarithmic correction only comes from two-form zero modes in in the asymptotically AdS$_4$ black hole background.  This result is essentially the same as in \cite{Bhattacharyya:2012ye}, however with the difference that here the 11d space is only locally $M_4\times S^7$, where $M_4$ is the AdS black hole.  (This difference manifests itself as $n_2^0=\chi_{\mathrm{AdS}}=1$ for global AdS$_4$ with $S^3$ boundary, in contrast to Eq.~(\ref{2form}) for the black hole.)  However, the Kaluza-Klein procedure, when performed properly, is equally valid in both cases.

The straightforward reduction of 11d supergravity on squashed $S^7$ does not yield any two-forms in four dimensions, as there are no non-trivial 1-cycles for the 11d three-form $C_{\mu\nu\rho}$ to be reduced on.  However, the quantization of $C_{\mu\nu\rho}$ introduces 2 two-form ghosts that are Grassmann odd, 3 one-form ghosts that are Grassmann even and 4 scalar ghosts that are Grassmann odd \cite{Siegel:1980jj}, and the two-form ghosts will contribute to the log term.

The 11d two-form ghost $A_2$ has action 
\be
S_2= \int A_2\wedge \star (\delta d+d \delta )^2A_2,
\ee
and the logarithmic term in the one-loop contribution to the entropy is thus, according to Eqs.~(\ref{entropy1lp})-(\ref{logterm}),
\be
\label{logL}
\log Z_{\text{1-loop}}[\beta,\ldots]=(2-\beta_2)n_2^0\log L+\cdots, 
\ee
where $\beta_2$ comes from integrating the zero modes in the path integral, and the minus sign takes care of the Grassmann odd nature of $A_2$.  The zero mode path integral becomes simply $\int [dA_2]|_{\text{zero modes}}$, and to find the logarithmic contribution in this term, one looks at the $L$ dependence by dimensional analysis, as in \cite{Bhattacharyya:2012ye}. The properly normalized measure is $\int d[A_{\mu\nu}]\exp(-L^7\int d^{11}x\sqrt{g^{(0)}}g^{(0)\mu\nu}g^{(0)\rho\sigma}A_{\mu\rho}A_{\nu\sigma})=1$, where we single out the $L$ dependence of the metric, $g_{\mu\nu}^{(0)}=\frac{1}{L^2}g_{\mu\nu}$. Thus the normalized measure is $\prod_x d(L^{\frac{7}{2}}A_{\mu\nu})$. For each zero mode, there is a $L^{\frac{7}{2}}$ factor. Thus in the logarithmic determinant, one has $\beta_2=\frac{7}{2}$. Combining  Eqs.~(\ref{2form}) and (\ref{logL}), the $\log L$ contribution to the thermal entropy in the extremal background is thus
\be
\label{partitionfunction}
\log Z_{\text{1-loop}}[\beta, \dots]=-3(1-g)\log L+\cdots.
\ee

{\it The extremal black hole entropy: } 
The coefficient of the logarithmic term in Eq.~(\ref{partitionfunction}) does not depend on $\beta$. In fact, due to the vanishing of the Seeley De-Witt coefficient, it can only depend on $\beta$ through regularized $n^0_p$'s, which, due to the asymptotic AdS condition, are topological. Therefore Eq.~(\ref{eq:entropy}) gives simply $S_{\text{1-loop}}=-3(1-g)\log L+\cdots$. As this is $\beta$ independent, it is also valid in the extremal limit, $\beta\to\infty$. Finally, the AdS/CFT dictionary establishes that $L\sim N^{1/6}$ leading to a logarithmic correction to the extremal black hole entropy of the form
\be
S_{\text{1-loop}}= -\frac{1-g}{2}\log N+\cdots,
\ee
which perfectly agrees with the microscopic result, (\ref{Eq:Index_g}).


{\bf Conclusions:} It is worth highlighting that the supergravity one-loop computation is universal in the sense that it applies to any asymptotically AdS$_4$ black hole that can be embedded in 11d supergravity under the mild condition that the seven-dimensional compactification manifold has vanishing first homology. There is a similar universal behavior in the one-loop effective action in AdS$_4$ \cite{Bhattacharyya:2012ye} which matches perfectly with the logarithmic correction of the supersymmetric partition function on $S^3$. It would be interested to establish the universality of the logarithmic corrections to the black hole entropy from the field theory side as well. 

Our precise example, when taken in conjunction with \cite{Liu:2017vll} and \cite{Jeon:2017aif}, clarifies that the quantum entropy function that has been so successful in the context of asymptotically flat black holes needs to be revisited in the context of asymptotically AdS black holes. Arguably, the connection between degrees of freedom residing at the horizon and other potential hair degrees of freedom needs to be better understood by revisiting  previous approaches  \cite{Banerjee:2009uk,Jatkar:2009yd}.

It was crucial in our result that we took a particular thermal-based  limit to the extremal black hole agreeing with some observations in the literature \cite{Sen:2008vm,Carroll:2009maa}.  This limiting procedure raises the specter that perhaps supersymmetric computations contain some information about slightly non-extremal systems in which case a window into capturing more dynamical information, such as Hawking radiation, could be opening.  


{\bf Acknowledgments:} We are thankful to A. Cabo-Bizet, I. Jeon, S. Lal, W. Fan, F. Larsen and especially to A. Sen for various explanations and encouragement. This work is partially supported by the US Department of Energy under Grant No.\ DE-SC0007859 and No.\ DE-SC0017808.


\bibliography{BHLocalization}

\end{document}